\documentclass[preprint,5p,twocolumn]{elsarticle}

\usepackage{amssymb}
\usepackage{amsmath}
\usepackage{hyperref}
\hypersetup{colorlinks=true,
            linkcolor=blue,
            filecolor=magenta,      
            urlcolor=cyan,
            pdftitle={An Example},
            pdfpagemode=FullScreen,}
\usepackage{tabularx}

\journal{Chemical Engineering Science}

\begin{document}
\begin{frontmatter}



\title{Temperature distribution in a gas-solid fixed bed probed by rapid magnetic resonance imaging}

\author[TUHH]{M. Raquel Serial}
\author[TUHH]{Stefan Benders}
\author[ETHZ1]{Perrine Rotzetter}
\author[TUHH]{Daniel L. Brummerloh}
\author[ETHZ1]{Jens P. Metzger}
\author[ETHZ2]{Simon P. Gross}
\author[ETHZ2]{Jennifer Nussbaum}
\author[ETHZ1]{Christoph R. Müller}
\author[ETHZ2]{Klaas P. Pruessmann}
\author[TUHH,ETHZ1,ETHZ2]{Alexander Penn}

\affiliation[TUHH]{organization={Institute of Process Imaging},
            addressline={Hamburg University of Technology}, 
            city={Hamburg},
            country={Germany}}

\affiliation[ETHZ1]{organization={Department of Mechanical and Process Engineering},
            addressline={ETH Zurich}, 
            city={Zurich},
            country={Switzerland}}
            
\affiliation[ETHZ2]{organization={Institute for Biomedical Engineering},
            addressline={ETH Zurich and University of Zurich}, 
            city={Zurich},            country={Switzerland}}

\begin{abstract}

Controlling the temperature distribution inside catalytic fixed bed reactors is crucial for yield optimization and process stability. Yet, in situ temperature measurements with spatial and temporal resolution are still challenging. In this work, we perform temperature measurements in a cylindrical fixed bed reactor by combining the capabilities of real-time magnetic resonance imaging (MRI) with the temperature-dependent proton resonance frequency (PRF) shift of water. Three-dimensional (3D) temperature maps are acquired while heating the bed from room temperature to 60~$^{\circ}$C using hot air. The obtained results show a clear temperature gradient along the axial and radial dimensions and agree with optical temperature probe measurements with an average error of $\pm$ 1.5~$^{\circ}$C. We believe that the MR thermometry methodology presented here opens new perspectives for the fundamental study of mass and heat transfer in gas-solid fixed beds and in the future might be extended to the study of reactive gas-solid systems.

\end{abstract}


\end{frontmatter}

\section{Introduction}
Temperature is a key parameter in reactive flow systems, affecting catalytic activity and thus conversion rates and reactor efficiencies. For instance, regions of above-average temperature (hot spots) can cause a reduced lifetime of catalysts or sorbents or can even lead to dangerous operating conditions \cite{hwang2008optimum}. Similarly, cold spots can result in a locally reduced catalytic activity or selectivity. 

The optimisation of heat transfer in reactors requires the understanding of how temperature gradients arise as a function of process parameters and reactor geometries. For this purpose, various measurement techniques have been proposed to probe the distribution of temperature. Traditional methods include thermocouples or optical fiber thermometers, which are commonly placed at different positions along the central axis \cite{huppmeier2010oxygen,graf2014experimental} or moved translatory through the bed using narrow capillaries \cite{horn2006syngas,wehinger2016investigating,geske2013resolving}. However, intrusive methods often provide single-point measurements and can cause disturbances to the gas flow pattern as well as temperature distortions due to thermal conduction of the sensor. Infrared radiation ~(IR)-thermography, on the other hand, can only provide spatially resolved temperature maps at the outside wall or surface of reactors if these are transparent to IR radiation~\cite{kopyscinski2010applying,kimmerle2009visualizing,simeone2008temperature}. 

Temperature measurements within full 3D volume of reactors call for the implementation of tomographic techniques. Laser-based methods such as near-infrared (NIR) tomography have been successfully applied to investigate the 3D distribution of temperature and species concentrations in fixed bed reactors \cite{alzahrani2019deactivation,an2012transient}. However, NIR-tomography studies have been, thus far, limited to low scattering media. Acoustic/ultrasonic tomography techniques are also capable of reconstructing temperature distributions utilizing the dependence of the speed of sound \cite{zhang2015online}. Yet, the employed signal reconstruction methods play a crucial role in obtaining reliable data from ultrasonic tomography \cite{liu2017reconstruction}.

Magnetic resonance imaging (MRI) is a tomographic technique that allows to quantify a variety of process parameters including, but not limited to, the local concentration of fluids, their velocity, diffusion coefficient and temperature \cite{hampel2022MRIreview}. Besides its widespread use in clinical practice and research, MRI is becoming increasingly important in the field of chemical engineering, owing in part to recent advances in temporal resolution \cite{penn2017real} that have enabled to study non-stationary flow systems including gas-solid fluidized beds \cite{penn2018fluidizedbeds, penn2019fb_obstacles} and spouted beds \cite{penn2020spout_regimes}.
Thermal contrast in MRI can be achieved by encoding any of the following temperature-dependent MR parameters \cite{WEBB20021}: the longitudinal relaxation time $T_\mathrm{1}$ \cite{parker1984applications,koptyug2008spatially,ridder2021spatially,skuntz2018melt}, the molecular self-diffusion coefficient \cite{le1989temperature,mirdrikvand2019diffusion}, the resonance frequency of a given chemical specie \cite{hindman1966proton,nott2005validation,raiford1979calibration,zhang2017accurate,silletta2019multinuclear} or the NMR linewidth of gases in the presence of a magnetic field gradient \cite{jarenwattananon2013thermal}. For example, the measurement of the difference in chemical shift of ethylene glycol (filled into spheres \cite{gladden2010mri} or narrow capillaries \cite{ulpts2015nmr}) has been demonstrated to provide accurate temperature measurements in the range of (37 – 165)~$^{\circ}$C without the need for any calibration \cite{van1968calibration,kaplan1975simplified,raiford1979calibration}. Among the temperature-dependent MR parameters that can produce thermal maps, proton ($^{1}$H) resonance frequency (PRF) \cite{ishihara1995precise} thermometry has been shown to be the most precise and sensitive method in detecting small thermal changes in fluids \cite{wlodarczyk1998three,wlodarczyk1999comparison}. Currently, PRF thermometry is widely used in interventional medical applications, such as high-intensity focused ultrasound (HIFU) \cite{holbrook2010real}, radiofrequency (RF) hyperthermia \cite{delannoy1990hyperthermia} and RF ablation \cite{van2008mri}. In the engineering field, PRF thermometry has been used recently to study convective heat transfer in heat exchangers and turbines \cite{bruschewski2021unbiased,buchenberg2016acquisition,bruschewski2022combined}.

In this work, we demonstrate the potential of PRF thermometry as an in-situ measuring technique to provide time-resolved, three-dimensional temperature measurements in a fixed bed reactor. Three-dimensional PRF experiments are conducted in a 70 mm wide model, fixed bed reactor during heating experiments at 60~$^{\circ}$C, with a temporal resolution of 4 seconds. The obtained results are validated against fiber optical sensors placed inside the bed during the MR acquisitions, showing a mean temperature deviation of $\pm$ 1.5~$^{\circ}$C. In addition, the measured temperature profiles were found to be in good agreement with numerical heat transfer simulations, demonstrating the capability of MRI-thermometry in probing the local temperature gradients inside fixed bed reactors. 

\section{Methods}\label{experimental_section}

\subsection{Fixed bed}
A cylindrical poly(methyl methacrylate)~(PMMA) bed of inner diameter of $D_\mathrm{bed} =$~\,70\,mm and height $H_\mathrm{bed} = $\,80\,mm was used for all experiments. The bed was filled with 282 hollow polypropylene~(PP) plastic spheres ($d_\mathrm{p,inner}\,=\,7.6\,$mm, $d_\mathrm{p,outer}\,=\,10\,$mm) containing a 36\,mM dysprosium-nitrate (Dy(III)(NO$_\mathrm{3}$)$_\mathrm{3}$) aqueous solution to mitigate image artifacts caused by the mismatch of the magnetic susceptibility between the diamagnetic spheres and the paramagnetic air surrounding them \cite{penn2018real, gross2016susceptometry}. The packing was heated by injecting 60\,$^{\circ}$C hot air via a distributor plate at the bottom of the bed featuring a central orifice of diameter of 5\,mm. A schematic representation of the system can be found in Fig.\ref{fig:scheme_bed}. A photograph is provided in the supplementary material (see Fig.~S1).
Gas flow was generated by a self-built 600\,W aluminum heating system, capable of heating over 180\,liters of air per minute (lpm) up to 100\,$^{\circ}$C. The outlet gas temperature was adjusted by means of a proportional-integral-derivative~(PID) controller. For all experiments, the temperature and flow rate of the gas leaving the heating unit were set to 60\,$^{\circ}$C and 100\,l/min at STP, respectively.

Both the fixed bed and the wind box were not insulated during the tests, so all the walls of the bed and the top and bottom of the bed were free to exchange heat with the environment.

\begin{figure}[ht!]
\centering
\includegraphics[width= \linewidth ]{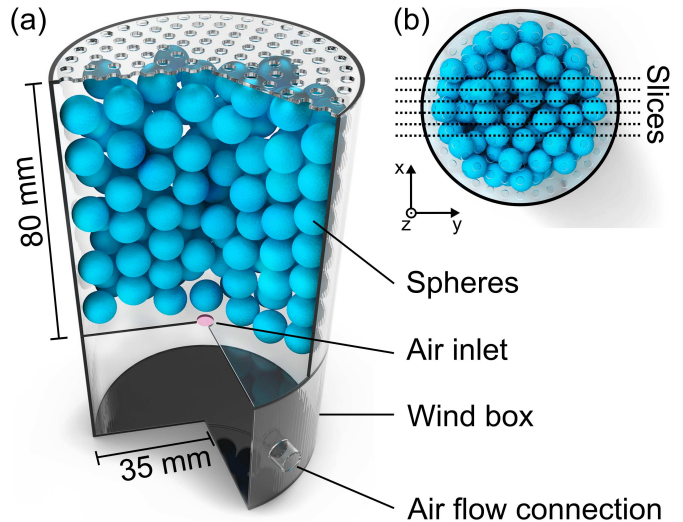}
\caption{\label{fig:scheme_bed} (a) Schematic representation of the fixed bed used for the MRI thermometry experiments. A PMMA plate with a single hole (5 mm diameter) is used as a gas distributor. (b) Top view of the bed showing the position of the 5 imaging slices.}
\end{figure}

\subsection{MRI thermometry: PRF-shift measurements}\label{PRF_method}

Proton resonance frequency~(PRF) shift experiments were conducted on a 7-T whole-body human MRI system (Achieva 7.0T, Philips Healthcare) using a 32-channel radio-frequency (rf) headcoil (Nova Medical, Inc.). Data processing was performed using MATLAB (Mathworks Inc, 2021b) equipped with the MRecon library (ReconFrame 3.0, GyroTools LLC, Zurich, Switzerland).

All MRI measurements reported here were acquired using a gradient echo~(GRE) pulse sequence with a temporal resolution of four seconds and a nominal spatial resolution of 1.01\,$\times$\,1.01\,mm$^2$. Five central (4\,mm wide) slices were acquired on each excitation, with a repetition time of 20\,ms, a flip angle of 25\,$^{\circ}$ and an echo time $t_\mathrm{Echo} = $ 1.75 ms.

The relative temperature fields at a time $t$ during heating experiments are calculated from the phase difference between a temperature and a reference scan:

\begin{flalign}\label{Tmap}
\Delta T(t) & = \frac{\phi(t) - \phi_{\mathrm{REF}}}{\alpha \gamma B_\mathrm{0} t_\mathrm{Echo}}&
\end{flalign}
where $\phi(t)$ and $\phi_{\mathrm{REF}}$ are the phase shift images of the temperature scan and reference scan, respectively, $\gamma$ is the gyromagnetic ratio of $^1$H, $B_\mathrm{0}$ is the magnetic field strength and $\alpha$ is the thermal coefficient of PRF-shift of the PP spheres \cite{Rieke2008376}. 

Absolute temperature fields ($T$) are obtained by adding the initial average temperature measured by three fiber-optical temperature sensors (Neoptix, Canada) to the measured relative temperature fields $\Delta T$. Each of these optical sensors were incorporated into the liquid filling of a sphere and sealed with silicon. These sensor spheres were placed along the centre axis of the bed at three different heights: bottom (20 mm), middle (42 mm) and top (61 mm).
For all experiments presented here, the first image during a heating experiment was taken as the reference to calculate the temperature maps. 

\subsubsection{Calculation of MRI thermal accuracy and precision}

The thermal accuracy and precision of MRI temperature measurements in a given region of interest (ROI) are calculated using the following equations:

\begin{flalign}\label{MRIaccuracy}
\begin{aligned}
\Delta T_\mathrm{accuracy} & = & \frac{1}{n}\sum_{t=t_1}^{t_n} \lvert \overline{T}_\mathrm{MRI,ROI}(t)  - T_\mathrm{sensors,ROI}(t)\rvert
\end{aligned}&&&
\end{flalign}

\begin{flalign}\label{MRIprecision}
\begin{aligned}
\Delta T_\mathrm{precision} & = &   \frac{1}{n}\sum_{t=t_1}^{t_n}  \mathrm{std}\left(\overline{T}_\mathrm{MRI,ROI}(t) \right)
\end{aligned}&&&
\end{flalign}
where $\overline{T}_\mathrm{MRI,ROI}(t)$ is the average MRI temperature within the ROI at time $t$, and $t_1$ and $t_n$ are the initial and final time steps measured, respectively and $n$ is the total time steps.

\subsection{Numerical simulations}\label{numerical_sim}

\subsubsection{Packing generation}

The numerical simulations were carried out in a 3D cylindrical 70\, mm diameter bed containing an spherical packing of spheres, resembling the geometry used for MRI experiments. A random packing of 282 hollow spheres was computed using the Bullet Physics library implemented in Blender 2.7 (Blender Foundation) employing the fixed bed generator (PBG) front end \cite{partopour2017integrated,partopour2020effect}. The generated spheres were nominally 10\,mm in diameter, with coefficients of friction and a restitution of 0.2 and 0.2, respectively. Contact points were avoided by shrinking the spheres by 2$\%$ of their diameters after their initialization.

\subsubsection{Heat transfer CFD simulations}

Transient heat transfer simulations were performed using the finite element method as implemented in COMSOL Multiphysics 5.5 employing the Heat Transfer in Solids and Fluids and Turbulent Flow $k-\epsilon$ modules.

Simulations were conducted on a 3D model consisting of a cylinder made of PMMA, with the same dimensions of the cylinder used for MRI experiments. The Blender generated packing was imported into COMSOL Multiphysics as a standard triangulation language (STL) file and discretised into tetrahedral mesh elements with a maximum spatial resolution of 12 elements per sphere diameter, as shown in Fig.~S2, and placed inside the cylinder. Bed spheres were assumed to be made of water, resembling MRI experiments, while the outer shell of the spheres was represented by a thin layer of 1.2 mm thickness made of PP using the COMSOL feature for layered materials. Overall dimensions of the simulated 3D model can be found in Table~\ref{table_sim_dimensions}.

\begin{table}
\begin{tabular}{lcl} \hline
& Value [mm] & Description \\ \hline
$D_\mathrm{bed}$ & 70 & Bed cylinder inner diameter\\ 
$D_\mathrm{bed,outer}$ & 80 & Bed cylinder outer diameter \\ 
$H_\mathrm{bed}$ & 80 & Bed cylinder height \\
$H_\mathrm{base}$ & 2 & Bed cylinder bottom base height \\
$D_\mathrm{inlet}$ & 5 & Bed inlet diameter \\ 
$d_\mathrm{p,inner}$ & 7.6 & Spheres inner diameter \\ 
$d_\mathrm{p,outer}$ & 10 & Spheres outer diameter \\ \hline
\end{tabular}
\caption{\label{table_sim_dimensions} Dimensions of the 3D model used for simulations}

\end{table}

Physical properties (such as density, viscosity, diffusivity, thermal conductivity and thermal capacity) related to the bed cylinder, gas phase and bed spheres were taken from the COMSOL Multiphysics material library and can be found in Table S1 in the supplementary material.  

Gas flow velocities (\textbf{u}) inside the bed were obtained by numerically solving the following equations, assuming a turbulent flow regime:

\begin{flalign}\label{Navier-Stokes_eq}
\begin{aligned}
\rho_g \frac{\partial \textbf{u}}{\partial t} + \rho_g(\textbf{u}\cdot \nabla)\textbf{u} & = \nabla \cdot [-p \textbf{I} + \textbf{K}] + \textbf{F}\\
\end{aligned}&&&
\end{flalign}

\begin{flalign}\label{continuity_eq}
&\frac{\partial\rho_g}{\partial t} + \nabla\cdot(\rho_g\textbf{u}) = 0&
\end{flalign}
where,
\begin{flalign}\label{K_eq}
\textbf{K}& = (\mu + \mu_T)(\nabla\textbf{u}+(\nabla\textbf{u})^T)- \frac{2}{3} (\mu + \mu_T)(\nabla\cdot\textbf{u})\textbf{I}-\frac{2}{3}\rho k \textbf{I}
\end{flalign}
and \textbf{u} and $\rho_g$ are the velocity and the density of the gas phase, respectively, $p$ is the static pressure, $\mu$ is the dynamic viscosity of the gas, \textbf{I} denotes the identity matrix and \textbf{F} is a volume force term.

The $k-\epsilon$ turbulence model \cite{wilcox1998turbulence} used for the simulations introduces two additional transport equations and two dependent variables: the turbulent kinetic energy ($k$) and turbulent dissipation rate ($\epsilon$), given by the following transport equations:

\begin{flalign}\label{kinetic_eq}
\rho_g \frac{\partial k}{\partial t} + \rho (\textbf{u}\cdot \nabla) k & = \nabla [(\mu + \frac{\mu_T}{\sigma_k}) \nabla k] + P_k -\rho_g \epsilon  
\end{flalign}
where the product term $P_k$ is
\begin{flalign}\label{Pk_eq}
P_k & = \mu_T [\nabla \textbf{u}:(\nabla \textbf{u} + (\nabla\textbf{u}^T)) - \frac{2}{3}(\nabla \cdot \textbf{u})^2] - \frac{2}{3} \rho k \nabla \cdot \textbf{u} 
\end{flalign}

The transport equation for $\epsilon$ reads:

\begin{flalign}\label{dissipation_eq}
\rho_g \frac{\partial \epsilon}{\partial t} + \rho (\textbf{u}\cdot \nabla) \epsilon & = \nabla [(\mu + \frac{\mu_T}{\sigma_{\epsilon}}) \nabla \epsilon] + \frac{C_{\epsilon_1}\epsilon}{k}p_k - \frac{C_{\epsilon_2}\rho \epsilon^2}{k}
\end{flalign}
and the turbulent viscosity is given by
\begin{flalign}\label{muT_eq}
\mu_T & = \rho C_{\mu} k^2/\epsilon 
\end{flalign}
where $C_{\epsilon_1}$, $C_{\epsilon_2}$ and $C_{\mu}$ are turbulence model constants, and $\sigma_k$ and $\sigma_{\epsilon}$ are the turbulent Prandl numbers for $k$ and $\epsilon$, respectively. All constants can be found in Table S2 in the supplementary material. 

Heat transfer in the numerical simulations included conduction and convection for the gaseous phase (Eq. \ref{Heat_transfer_gas_eq}) and conduction for the spheres phase (Eq. \ref{Heat_transfer_solid_eq})

\begin{flalign}\label{Heat_transfer_gas_eq}
\begin{aligned}
\rho_g C_{pg} \frac{\partial T}{\partial t} + \nabla \cdot (-k_g \nabla T) + \rho_g C_{pg} \textbf{u} \cdot \nabla T & = 0 &
\end{aligned}&&&
\end{flalign}

\begin{flalign}\label{Heat_transfer_solid_eq}
\begin{aligned}
\rho_s C_{ps} \frac{\partial T}{\partial t} + \nabla \cdot (-k_s \nabla T) = 0 &
\end{aligned}&&&
\end{flalign}
where $C_{pg}$ and $C_{ps}$ are the heat capacity of the gas and spheres phases, respectively, and $k_g$ and $k_s$ are the thermal conductivity of the gaseous and spheres phases, respectively. Heat conduction between the core and the shell of the bed spheres is also solved during the simulations. 

Boundary conditions used for the bed cylinder, the gas phase and the bed spheres can be found in Table S3, S4 and S5 respectively in the supplementary material.

\section{Results and discussion}

\begin{figure}[!ht]
\centering
\includegraphics[width= 7.0 cm]{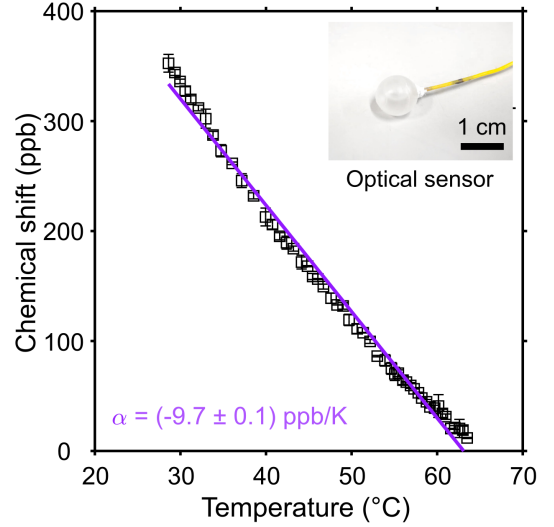}
\caption{\label{fig:calibration} ~Temperature dependence of the chemical shift for a single PP particle filled with 36\,mM Dy(III)(NO$_\mathrm{3}$)$_\mathrm{3}$ aqueous solution. The PRF-shift coefficient ($\alpha$) is determined from the slope of a linear fit (purple line). Temperature was monitored by means of a fiber-optical temperature sensor placed inside the PP shell (inset). Symbols refer to means over two replicates, and error bars represent the deviation range.}
\end{figure}

First, the PRF-shift coefficient ($\alpha$) of the bed spheres is determined. Figure~\ref{fig:calibration} shows the chemical shift temperature dependence of a 10 mm diameter PP hollow sphere filled with a 36\,mM Dy(III)(NO$_\mathrm{3}$)$_\mathrm{3}$ aqueous solution. For this calibration measurement, a single sphere was first heated to 65 $^{\circ}$C and subsequently cooled down to room temperature inside the bed. During the cooldown period, the temperature is monitored by a fiber optic temperature sensor placed inside the sphere (see Fig.~\ref{fig:calibration} inset) while the PRF shift is recorded using the MR thermometry sequence. 

As expected, the measured chemical-shift values exhibited a linear trend with decreasing temperature, although a slight curvature is observed for temperatures above 50 $^{\circ}$C. The source of this nonlinearity is the complex thermal dependence of the intermolecular forces between water molecules and has been described previously \cite{hindman1966proton}. For the temperature range studied in our work, we consider a linear fit to be a reasonable choice, although higher temperatures would require a different calibration approach such as a lookup calibration table. The PRF-shift coefficient for the bed spheres is then determined from the slope of a linear fit, giving $\alpha$ = (-9.7\,$\pm$\,0.1)\,ppb/$^{\circ}$C, consistent with $\sim$\,-10\,ppb/$^{\circ}$C as reported in previous studies \cite{hindman1966proton,ishihara1995precise,mcdannold2005quantitative}.

\begin{figure*}[ht]
\centering
\includegraphics[width=\textwidth]{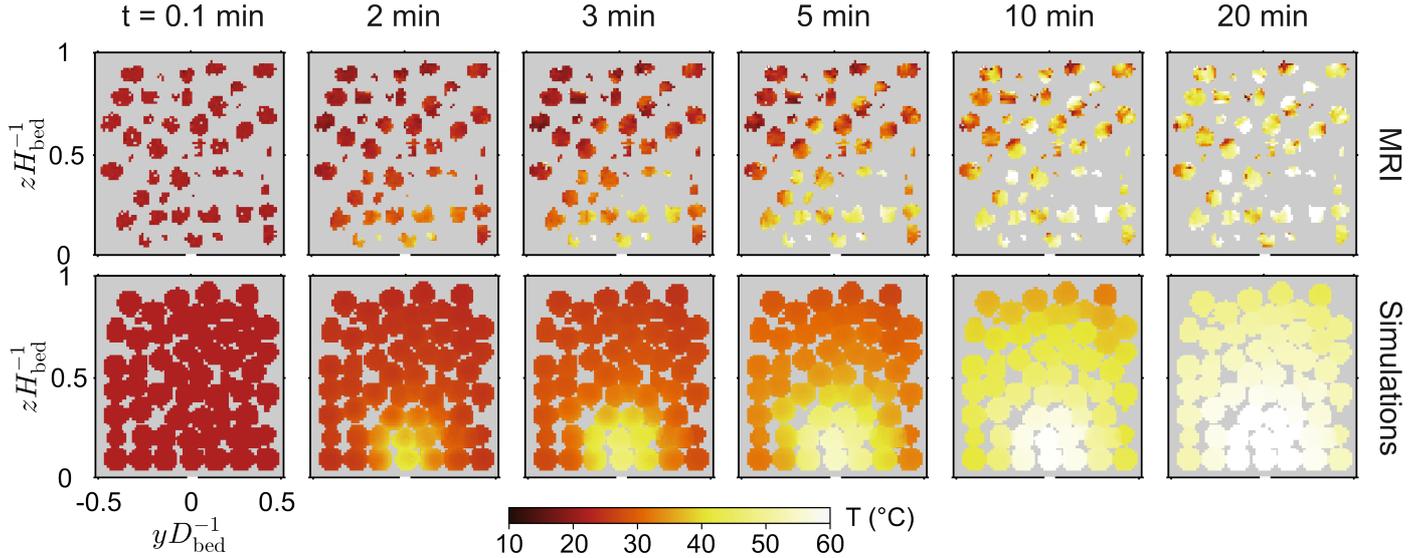}
\caption{\label{fig:Tmap_heating} MRI (top) and simulated (bottom) temperature fields of a fixed bed during gas flow experiments (100 l/min) at 60\,$^{\circ}$C. Temperature maps are shown for a central slice (slice 3) of 4\,mm width. Open lines at the bottom of the plots show the position of the gas injection port. MRI spatial resolution: 1.01\,$\times$\,1.01\,mm$^2$, MRI temporal resolution: 4\,s.  A similar behavior is obtained for the recorded five slices (see video 1 in the supplementary information).}
\end{figure*}

Following the temperature calibration, PRF-shift thermometry measurements were performed in a fixed bed reactor (Fig.\ref{fig:scheme_bed}) while heating the packing from room temperature by means of a 60\,$^{\circ}$C hot airflow.  Figure~\ref{fig:Tmap_heating} shows time series of the measured MRI temperature fields for a single central slice (slice 3) of the 3D packing. Absolute temperature maps were obtained by adding the initial sample temperature, measured by optical sensors placed inside the bed. As expected, the bed initially shows a fairly uniform temperature distribution ($T\sim$ 23\,$^{\circ}$C). As time increases, temperature heterogeneities can be observed: spheres closest to the gas inlet start heating up, while the temperature at the top of the bed remains nearly unchanged. After only five minutes, the entire bed has increased its temperature displaying a gradient along the axial ($z$) and radial ($r$) directions. Finally, after 20 minutes the temperature is fairly homogeneous, reaching values $>\,50\,^{\circ}$C for most spheres within the bed. A similar behavior is obtained for the recorded five slices (see video 1 in the supplementary information). 

It is important to note that thermal inhomogeneities within a single sphere are visible at different points of the bed in Fig. 3. These fluctuations occur mainly at the edges of the spheres and increase with the duration of heating. Mentioned variations could be related to the softening and subsequent deformation of the PP spherical shells with temperature. The latter behaviour can be observed clearly in the subtraction of the first and last image taken during heat flow experiments (Fig.~S3 in the supplementary material), where changes at the edges of the spheres are visible. These few pixels deformation, even though were not considered for the calculation of temperature maps, results in magnetic susceptibility field differences that cannot be corrected with a reference image after a certain time. This is an inherent limitation of our approach, as it is based on the assumption of constant magnetic susceptibility with temperature. In the future, this could be improved by using particles that do not deform as easily with temperature. 

\subsection{Validation of temperature measurements}

\begin{figure}[!ht]
\includegraphics[width= 8.5 cm]{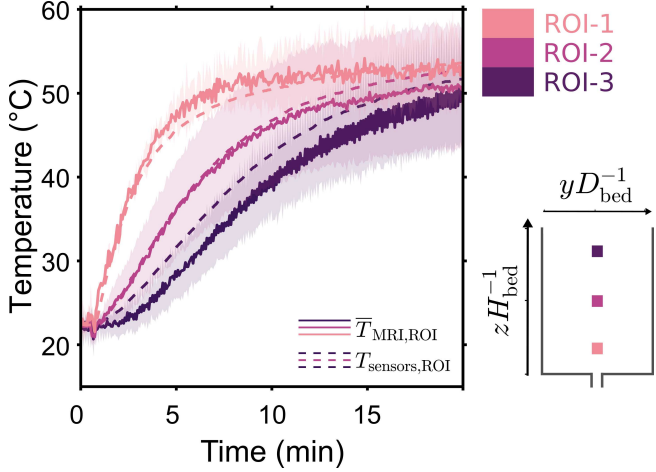}
\caption{\label{fig:MRI_vs_sensors} Comparison between MRI PRF-shift measurements (solid lines) and fiber optic temperature readings (dashed lines) during gas flow at 60\,$^{\circ}$C, at three different locations along the bed. MRI temperature values at the three selected ROIs of dimensions 2\,$\times$\,2\,$\times$\,20\,mm$^3$ were averaged.}
\end{figure}

The accuracy of the MRI temperature measurements is assessed by comparison with readings of optical temperature sensors, placed at different heights within the bed. Figure~\ref{fig:MRI_vs_sensors} shows the position of three regions of interest~(ROI) of dimensions (20\,$\times$\,2\,$\times$\,2)\,mm$^3$ along the $x$, $y$ and $z$-directions which are used to monitor the temperature values at the bottom~(ROI$_\mathrm{1}$), medium~(ROI$_\mathrm{2}$) and top~(ROI$_\mathrm{3}$) of the bed along the central vertical axis ($y = 0$) of the reactor model. The mean MRI temperature values within the three ROIs ($\overline{T}_\mathrm{MRI,ROI}$) are plotted as solid lines in Figure~\ref{fig:MRI_vs_sensors} along with the fiber optic probe measurements ($T_\mathrm{sensors,ROI}$) given as dashed lines. The corresponding standard deviations of calculated $\overline{T}_\mathrm{MRI,ROI}$ are shown as shaded areas.
It is worth to mention that all 5 slices are considered in the averaged MRI temperature values. 

Figure~\ref{fig:MRI_vs_sensors} shows that both, the sensors readings and MRI-based temperatures, display a clear temperature gradient along the axial direction for heating times $t<$~10\,min. Once the gas flow is initiated, the ROI$_\mathrm{1}$ temperature exhibits a rapid increase, reaching $T\sim$\,48\,$^{\circ}$C after only 5 minutes. The middle (ROI$_\mathrm{2}$) and top (ROI$_\mathrm{3}$) of the bed exhibit a similar trend, but at a slower rate. A comparable temperature-time evolution was reported by several authors when studying transient temperature gradients in a fixed bed reactor, employing thermocouples \cite{wen2006heat,messai2014experimental}.

We note that the final bed temperatures as measured by the optical sensors and the proposed MRI method, deviate from the reference gas temperature (60~$^{\circ}$C). The latter is measured at a distance of 100\,mm from the bed (see yellow connection in Fig.~S1). The discrepancies observed can be attributed to a lack of insulation at the wind box walls, resulting in a slightly lower temperature of the entering gas and thus lower bed temperatures.

The comparison between MRI and optical temperature measurements shows a good agreement for the entire time range studied and an average thermal accuracy of 1.5\,$^{\circ}$C for the three ROIs, calculated using Eq.~\ref{MRIaccuracy}. Thermal deviations of $\overline{T}_\mathrm{MRI,ROI}$ and sensor probe measurements for each ROI as a function of time can be observed in Fig.~S4 in the supplementary material, where higher deviations are observed at the top of the bed.  
Note that while the average MRI temperatures are consistent with the optical sensor readings, there is considerable thermal scatter within each ROI as observed in Fig.~\ref{fig:MRI_vs_sensors}. The latter results in an average thermal accuracy of 4.4~$^{\circ}$C for the three ROIs, calculated according to Eq.~\ref{MRIprecision}. 

\subsection{Analysis of temperature heterogeneities}

To test whether MRI-thermometry can successfully capture temperature gradients, we compare the obtained MRI temperature maps with numerical heat transfer simulations. Bottom panel in Fig. \ref{fig:Tmap_heating} shows the time evolution of simulated temperature maps for an inlet temperature of 60\,$^{\circ}$C. We find that the measured temperature fields are in good agreement with predicted temperature maps, although more pronounced radial temperature gradients are observed in the simulation results. 

\begin{figure*}[!ht]
\centering
\includegraphics[width= 12 cm]{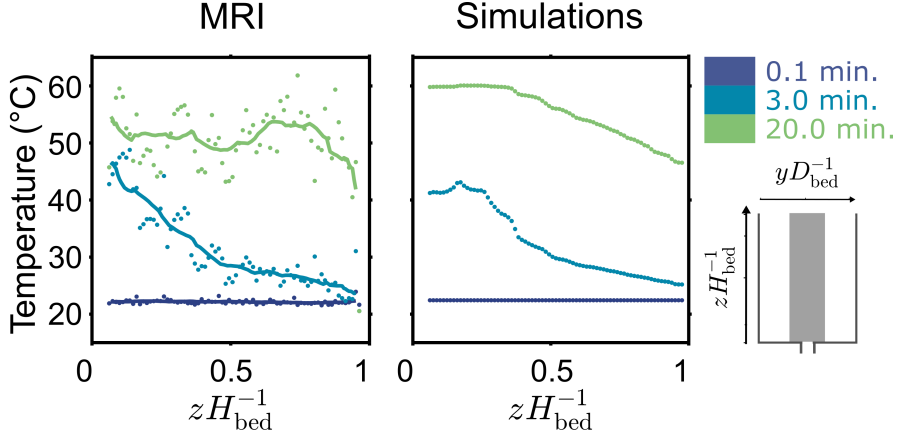}
\caption{\label{fig:Z_profiles_vs_time} Averaged axial temperature profiles as obtained by MRI and numerical simulations at ~$y/D_\mathrm{bed}\,=$\,0 for three different times during heating experiments. Solid lines in the MRI temperature plots represents a moving average filter with a span window of 18 points. The averaged region (20\,mm\,$\times$\,20\,mm\,$\times$\,80\,mm) along the $x$, $y$ and $z$ direction is schematically represented in grey.}
\end{figure*}

Spatial temperature heterogeneities can be analysed in further detail by assessing 1D temperature profiles at different positions along the bed and comparing MRI results to numerical heat transfer simulations. Figure \ref{fig:Z_profiles_vs_time} shows the resulting MRI and simulated temperature profiles along the axial dimension for three different times during the heating experiments. The averaged volume (20 mm\,$\times$\,20 mm\,$\times$\,80 mm) along the $x$, $y$ and $z$ direction is schematically represented in grey. At short times ($t\leq$ 3 minutes), MRI temperature profiles reveal a clear temperature gradient along the axial dimension. The latter is the result of a temperature front propagating across the bed. While the particles at the bottom heat up first, they reduce the temperature of the gas at the front, which leads to a delay in the heating of the particles further up the bed. This behaviour is influenced by the gas velocity field at the bed entrance, as previously reported by \cite{donaubauer20192d}. The observed axial temperature gradient in Fig.\ref{fig:Z_profiles_vs_time} equilibrates after about 20 minutes of heating. Numerical simulations of heat transfer show similar spatial temperature variations, although significant differences are observed after 20 minutes of heating, where a temperature gradient is still observed. Found differences could be related to an inaccurate velocity field within the fixed bed, caused by the shrinking of the spheres diameter to avoid meshing problems.

\begin{figure*}[!ht]
\centering
\includegraphics[width= 12 cm]{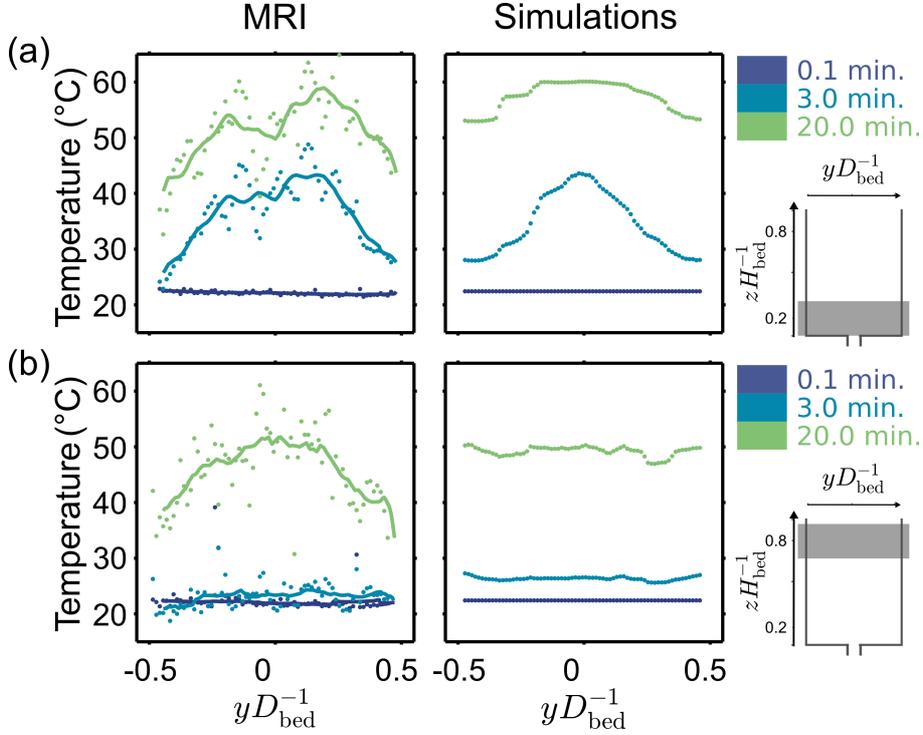}
\caption{\label{fig:R_profiles_vs_time} MRI and numerical simulations based, averaged radial temperature profiles for (a)~$z/H_\mathrm{bed}\,=$\,0.2 and (b)~$z/H_\mathrm{bed}\,=$\,0.8 at three different times during a heating experiment. Solid lines in MRI temperature plots represents a moving average filter with a window of 7 points. The averaged region (20\,mm\,$\times$\,70\,mm\,$\times$\,20\,mm) along the $x$, $y$ and $z$ direction is highlighted in grey.}
\end{figure*}

Furthermore, we study the radial temperature heterogeneities. Figure \ref{fig:R_profiles_vs_time} depicts MRI and simulated radial temperature profiles at $z/H_\mathrm{bed}\,=$\,0.2 (a) and $z/H_\mathrm{bed}\,=$\,0.8 (b) at three different stages during the heating experiments. Temperatures for both MRI and simulated data sets were averaged along the five recorded slices. MRI profiles exhibit a clear radial dependence of temperature at both the bottom and the top of the bed. This temperature gradient builds up soon after the start of the heating and persists for the duration of the experiment.  After 20 minutes, the temperature at the bottom of the bed becomes slightly more homogeneous, although a radial dependence is maintained at both positions. 

Heat transfer simulations (right panel in Fig.\ref{fig:R_profiles_vs_time}) were found to be in good agreement with the measured MRI temperature profiles, displaying temperature values in the same numerical range, although slightly higher temperatures are found after 20 minutes of heating. 
Higher deviations are observed in the regions close to the bed wall, where MRI temperatures decrease notoriously. The observed discrepancies could be due to an inaccurate modelling of the shell and core structure of the spheres, which affects not only the heat conduction between the spheres but also between the packing and the bed walls. Furthermore, changes in the contact area between the spheres, e.g. by reducing the sphere diameter (2~\%), as was done in this work to avoid meshing problems, can also affect the radial heat transfer behaviour \cite{wehinger2017contact}.
It is interesting to note that both the experimental and simulated profiles are slightly asymmetric, suggesting that this effect is more influenced by the arrangement of the packing.

\subsection{Derivation of effective heat transfer parameters}

Evaluation of temperature variations shown in Figure~\ref{fig:MRI_vs_sensors} suggests that the time-dependent temperature of the bed spheres can be described by a simplification of the Schumann’s model \cite{schumann1929heat} given by the following energy equation for a volume voxel-$i$: 

\begin{flalign}\label{Schaumann_eq}
\begin{aligned}
\rho_s C_{ps}\frac{\partial T_i}{\partial t} = \frac{6}{d_p}  h_i \Delta T_i
\end{aligned}&&&
\end{flalign}
where $\rho_s$ and $C_{ps}$ are the density and heat capacity of the solid phase, respectively, $d_p$ is the diameter of the bed spheres and $T_i$, $\Delta T_i$ and $h_i$ are the temperature, temperature difference and heat transfer coefficient associated to the voxel $i$, respectively. Solving Eq.\ref{Schaumann_eq} for the initial, $T_{i,t=0}$, and final, $T_{i,t=\infty}$, voxel temperatures, gives:

\begin{flalign}\label{exp_build_up}
\begin{aligned}
T_i(t) = (T_{i,t=\infty} - T_{i,t=0}) \left[1-\exp\left(-\frac{6 h_i t}{\rho_{s} C_\mathrm{ps} d_p}\right)\right] + T_{i,t=0}
\end{aligned}&&&
\end{flalign}

Fitting of Eq.~\ref{exp_build_up} for each volume pixel of acquired MRI temperature maps, enables the determination of a local heat transfer coefficient $h(r,z)$ for a given radial $r$ and axial $z$ position. The volume pixels positions were transformed into cylindrical coordinates in order to allow an accurate mapping of the volume pixels across the slices. Least squares fitting did not reveal any significant differences in the heat transfer along the radial direction (see Fig.~S5). Therefore, the obtained heat transfer coefficients were averaged along the radial direction.

\begin{figure}[!ht]
\centering
\includegraphics[width=8.5 cm]{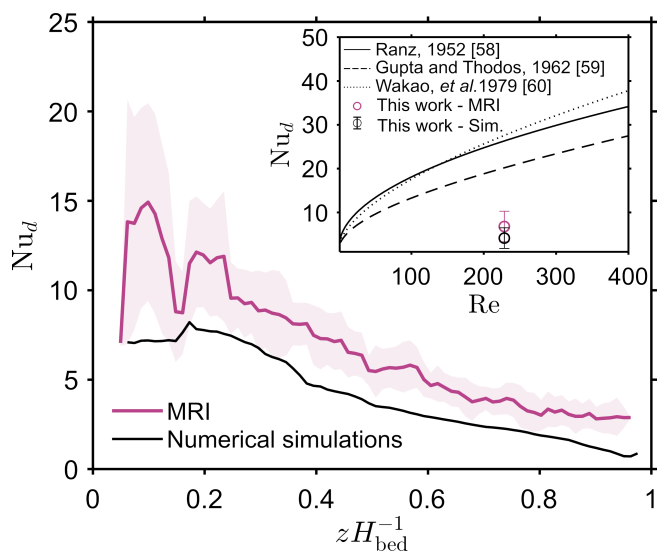}
\caption{\label{fig:Nusselt_MRI_vs_COMSOL} MRI and simulated local Nusselt numbers ($Nu_d$) as a function of the normalised axial position. Shaded areas correspond to MRI standard deviations along the radial direction. Inset: $Nu_d$ versus Reynolds number correlations for references \cite{ranz1952friction} (solid line), \cite{gupta1962mass} (dashed line) and \cite{wakao1979effect} (dotted line) and the average $Nu_d$ and standard deviation (error bar) calculated from MRI and simulated data.}
\end{figure}

In order to determine the heat transfer properties of the bed, we focus on the position dependence of the dimensionless Nusselt number Nu$_d$. The Nu$_d$ number gives the ratio between convective and conductive heat transport. The Nusselt number (Nu$_d = h(z)d_p/k_{g}$) calculated from the fitted heat transfer coefficient $h(z)$ and the thermal conductivity of the gas (air) $k_g$, is shown in Fig.~\ref{fig:Nusselt_MRI_vs_COMSOL} as a function of the normalised axial position. For comparison, the local Nusselt numbers fitted from numerical heat transfer simulations are plotted as solid lines in the same graph. Both, MRI and the simulated data display a decrease in local Nusselt number with increasing distance from the gas inlet. Although the profiles show a remarkably similar trend, the simulated local Nusselt number is considerable smaller along the entire bed axis. Once again, found differences could be related to the approximations and assumptions made for the 3D model mentioned in previous sections, such as the modelling of the spheres shell as well as the shrinking of its diameter.

Several experimental Nusselt number correlations are available in the literature \cite{esence2017review}. Inset in Fig.~\ref{fig:Nusselt_MRI_vs_COMSOL} shows a comparison of the average MRI obtained Nusselt number with correlations given by Ranz \textit{et al.} \cite{ranz1952friction}, Gupta and Theodos \textit{et al.} \cite{gupta1962mass} and Wakao \textit{et al.} \cite{wakao1979effect}. It can be seen that even though the presented results are in the same order than correlations found in previous literature, considerable discrepancies are found when different bed particles, setup dimensions and boundary conditions are considered, as previously reported by Messai \textit{et al.} \cite{messai2014experimental}. In fact, most experiments employed for the calculation of correlations, have considered adiabatic conditions for the bed walls which differs substantially with the experiments carried out in this work. In addition, the material properties of the bed particles particularly differ from the commonly used particles in fixed beds heat transfer experiments, where commonly solid particles are used.

The work presented here uses the water contained in plastic shells as a temperature probe. The fact that water has a higher heat capacity compared to most solid materials, might result in a slower thermal response as compared to a bed of solid particles. In the future, the aqueous solution could be replaced by a different liquid that exhibits a temperature-dependent PRF-shift and that has a higher boiling point and lower heat capacity as compared to water (e.g. ethylene glycol). These modifications would increase the applicability of the technique presented here to reactive gas-solid systems. 
\section{Conclusions}
In this paper, we demonstrated the capabilities of MRI-thermometry in obtaining spatially and temporally resolved, quantitative heat transfer measurements within a 3D fixed bed during heating experiments. The proposed methodology provided detailed temperature maps with a temporal resolution of four seconds, enabling to resolve temperature heterogeneities across the bed. The results obtained were found to be in good agreement with fiber optic probes placed inside the bed, as well as with transient heat transfer numerical simulations. Finally, MRI temperature maps were used to determine the heat transfer coefficient of the bed spheres. The latter was found to be in the same order of magnitude reported in previous works. We anticipate that the methodology presented in this work might be used to study more complex reactor models including different bed and particle geometries, gas distribution systems, and reactive systems.

\section{Declaration of Competing Interest}
The authors declare that they do not have any competing interests.
\section{Author contributions}
A.P., K.P.P., S.P.G., and C.R.M. conceived and designed the analysis. A.P. and S.P.G. designed the MR pulse sequence. P.R. and A.P. designed and built the experimental setup and collected the data. J.N. collected the optical reference temperature measurements. M.R.S, S.B., D.L.B. and J.P.M. performed the numerical simulations. M.R.S., S.B. and A.P. analysed the data. M.R.S., S.B., A.P., J.P.M., and C.R.M. wrote the paper.

\section{Acknowledgements}
This work was supported by the Swiss National Science Foundation under the grant no. 200020$\_$182692. 
\bibliographystyle{elsarticle-num}
\bibliography{cas-refs}

\end{document}